\documentclass[twocolumn,aps,pra,showpacs,groupedaddress,superscriptaddress,nofootinbib]{revtex4-1}

\usepackage[pdftex,linkcolor=blue,citecolor=blue,urlcolor=blue,colorlinks]{hyperref}
\usepackage[dvipsnames]{xcolor}
\usepackage{graphicx,epsfig}
\usepackage{braket}
\usepackage{color}
\usepackage{cancel}
\usepackage{amsmath}
\usepackage{amsfonts}
\usepackage{fancyhdr}

\begin{document}
\title{Quantum magnetism with ultracold bosons carrying orbital angular momentum}
\author{G. Pelegr\'i}
\affiliation{Departament de F\'isica, Universitat Aut\`onoma de Barcelona, E-08193 Bellaterra, Spain.}
\author{J. Mompart}
\affiliation{Departament de F\'isica, Universitat Aut\`onoma de Barcelona, E-08193 Bellaterra, Spain.}
\author{V. Ahufinger}
\affiliation{Departament de F\'isica, Universitat Aut\`onoma de Barcelona, E-08193 Bellaterra, Spain.}
\author{A. J. Daley}
\affiliation{Department of Physics and SUPA, University of Strathclyde, Glasgow G4 0NG, United Kingdom.}

\begin{abstract}
We show how strongly correlated ultracold bosonic atoms loaded in specific orbital angular momentum states of arrays of cylindrically symmetric potentials can realize a variety of spin-1/2 models of quantum magnetism. We consider explicitly the dependence of the effective couplings on the geometry of the system and demonstrate that several models of interest related to a general $XYZ$ Heisenberg model with external field can be obtained.   
Furthermore, we discuss how the relative strength of the effective couplings can be tuned and which phases can be explored by doing so in realistic setups. Finally, we address questions concerning the experimental read-out and implementation and we argue that the stability of the system can be enhanced by using ring-shaped trapping potentials.

\end{abstract}
\maketitle


\section{Introduction}
Ultracold atoms in optical lattices provide a clean and highly tunable playground to study a plethora of many-body phenomena \cite{bookVeronica}. Recent years have witnessed important breakthroughs that have pushed the degree of control over these systems to a very precise quantitative level and have opened new routes towards the quantum simulation of previously unexplored systems in a wide range of fields \cite{reviewOptLattices}. In particular, ultracold atoms have proven to be a very powerful tool for exploring quantum magnetism in a form originally inspired by solid state systems. Remarkable achievements of quantum simulation of magnetism with ultracold atoms include the implementation of spin-frustrated lattices \cite{FrustratedMagnetismBosons1, FrustratedMagnetismBosons2}, extensive experimental studies of the magnetic properties of the Hubbard model \cite{MagnetismFermions1,MagnetismFermions2,MagnetismFermions3,MagnetismFermions4,
MagnetismFermions5,MagnetismFermions6,MagnetismFermions7,MagnetismFermions8}, or the realization of high-resolution quantum gas microscopes for bosonic atoms \citep{QGasMicro1,QGasMicro2} that have lead to the observation of anti-ferromagnetic order in a one-dimensional (1D) Ising chain \cite{1DIsingBosons}, bound magnons in the $XXZ$ Heisenberg model \cite{MagnonsBosons} and spin-resolved dynamics \cite{SpinImpurityBosons1, SpinImpurityBosons2, SpinImpurityBosons3}. There are also proposals to realize spin models with strongly interacting ultracold bosons excited to $p-$bands \cite{XYZpinheiro}, and realizations of magnetic models with bosons in tilted optical lattices \cite{1DIsingBosons,MottEfield,SpinTiltedLattice,QuenchTiltedLattice1,QuenchTiltedLattice2}.    

In this paper, we show that strongly interacting ultracold bosons loaded into Orbital Angular Momentum (OAM) states of lattices of side-coupled cylindrically symmetric traps can realize a variety of spin-$1/2$ models, including the $XYZ$ Heisenberg model with or without external field. In particular, we focus on the Mott insulator regime at unit filling, where each trap is occupied by a single atom and a direct mapping between the OAM and spin-$1/2$ states can be performed. Recently, a proposal to realize such a state by periodically modulating an optical lattice has been made \cite{MugaOAM}. Alternatively, this state could be generated by optically transferring OAM \cite{OptOAMAtoms} to atoms confined to an arrangement of ring-shaped potentials, which can be created by a variety of techniques \cite{ring1,ring2,ring3,ring4,ring5,ring6,ring7,DoubleRing,TAAP,TimeAveragedBECRing,TimeAveragedBECRing2} and have proven to support long-lived persistent currents associated to the OAM states \cite{persistent1,persistent2}. The mechanisms that yield these effective spin-$1/2$ models are analogous to the ones described in \cite{XYZpinheiro}, where it was shown that the $XYZ$ Heisenberg model can be realized with ultracold bosons in the $p$-bands of a two-dimensional optical lattice \cite{HigherOrbital1,HigherOrbital2}, which are equivalent to the OAM $l=1$ states. Our proposal, however, extends this to lattices made up of general cylindrically symmetric potentials such as ring traps, and is valid for higher OAM states. The new degree of control offered by the flexibility in the arrangements of the traps opens up the possibility to engineer a wide variety of spin models beyond the $XYZ$ Heisenberg model and makes it possible to modify the effective coupling parameters at the level of a single site. 

The rest of the paper is organized as follows. In section \ref{physical}, we describe the general physical system and give details of how to compute the couplings that govern the effective spin-$1/2$ model. In section \ref{models}, we make concrete proposals to implement different spin-$1/2$ models of interest by arranging the ring potentials in different geometries. In section \ref{control}, we discuss how the effective couplings can be tuned experimentally and which phases of the $XYZ$ model can be explored by doing so. In section \ref{experimental}, we discuss the implementation of lattices of ring potentials and the readout and stability of OAM states in an experimental realization. Finally, in section \ref{conclusions} we summarize the main conclusions of this work.
\section{Quasi one-dimensional ladder and effective spin-$1/2$ model}
\label{physical}
For the sake of clarity, we start by considering in this section the simplest quasi one-dimensional lattice in which an effective spin-$1/2$ model of quantum magnetism can be obtained using ultracold atoms carrying OAM, namely an array of equivalent ring-shaped potentials. From the analysis of the second order processes that we will discuss for this system, the generalization of the effective spin model to other quasi one-dimensional geometries and to two-dimensional lattices is straightforward.

\begin{figure}[t!]
\centering
\includegraphics[width=\linewidth]{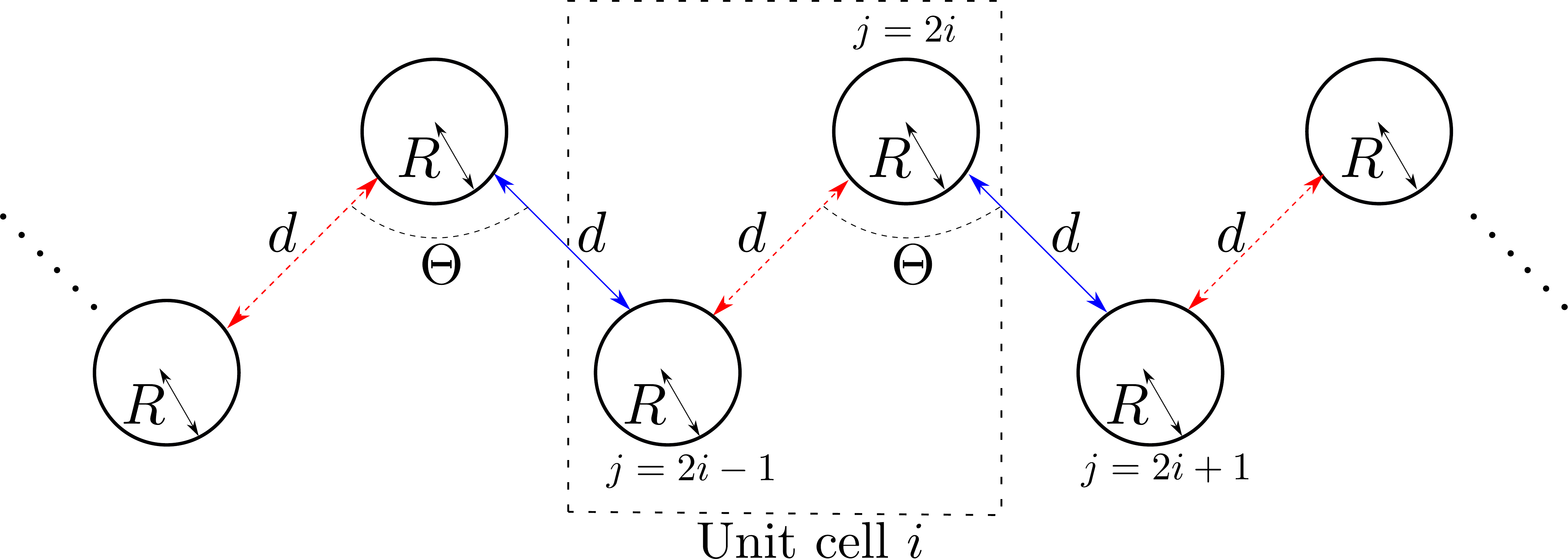}
\caption{Quasi one-dimensional ladder of ring potentials (labelled by the index $j$) obtained by concatenating unit cells (labelled by the index $i$) formed by two rings such that the central angle of the triangles formed by three neighbouring rings is $\Theta$. The origin of phases is taken along the direction $2i\leftrightarrow 2i+1$ (indicated with blue straight arrows), so that the couplings are real along this direction and all hopping phases appear in the $2i\leftrightarrow 2i-1$ links (indicated with red dashed arrows). The distance between the closest points of two nearest-neighbour rings is $d$.}
\label{PhySystem}
\end{figure}

The quasi one-dimensional system on which we focus consists of a gas of $M$ ultracold bosons of mass $m$ trapped in a ladder of $N$ identical ring-shaped potentials, labelled by the index $j$. This can be constructed by concatenating $N/2$ two-ring unit cells, labelled by the index $i$, as depicted in Fig. \ref{PhySystem}. All of the rings have the same radius $R$ and radial trapping frequency $\omega$, which defines the natural length scale $\sigma=\sqrt{\hbar/m\omega}$. The outer parts of two rings belonging to the same unit cell are separated a distance $d$, and three consecutive rings form a triangle with a central angle $\Theta$. The bosons may occupy the two degenerate eigenstates of total OAM $l\geq 1$ of each ring, $\ket{j,\pm l}$, for which the wavefunctions are given by
\begin{equation}
\phi_{\pm l}^{j}(r_{j},\varphi_{j})=\braket{\vec{r}|j,\pm l}=\psi_l(r_j)e^{\pm i l(\varphi_j-\varphi_0)},
\label{wavefunctions}
\end{equation}
where $(r_j,\varphi_j)$ are the polar coordinates with origin at the center of the $j$th ring and $\varphi_0$ is an arbitrary origin of phases. The radial part of the wavefunction, $\psi_l(r_j)$, can be approximated by the ground state of the $j$th ring potential, $\psi_0(r_j)$. Under this approximation, the energy of the modes of OAM $l$ is given by
\begin{equation}
E(l)=E_0+E_{\text{c}}l^2,
\label{energiesOAM}
\end{equation}
where $E_0$ is the energy of the ground state of the ring and $E_{\text{c}}=\frac{\hbar^2}{2m}\int d^2r\left|\frac{\psi_0(r)}{r}\right|^2$ is the centrifugal part of the kinetic energy.  
We assume that the motion of the bosons is restricted to the manifold of states of total OAM $l$, i.e., values of $\pm l$, without coupling to other OAM manifolds. In this situation, the total bosonic field operator of the system reads  
\begin{equation}
\hat{\Psi}_l=
\sum_{j=1}^{N} \phi_{+l}^j(r_j,\varphi_j)\hat{a}_{+ l}^{j}+\phi_{-l}^j(r_j,\varphi_j)\hat{a}_{- l}^{j}
\label{bosonicfield}
\end{equation}
where $\hat{a}_{\pm l}^{}$ are the bosonic annihilation operators associated with the respective OAM modes.
The Hamiltonian of the system can be decomposed into its single particle and interacting parts
\begin{align}
\hat{H}_l&=\int d\vec{r}\hat{\Psi}_l^{\dagger}\left[-\frac{\hbar^2\nabla^2}{2m}+V(\vec{r})\right]\hat{\Psi}_l
+\frac{g}{2}\int d\vec{r} \hat{\Psi}_l^{\dagger}\hat{\Psi}_l^{\dagger}\hat{\Psi}_l\hat{\Psi}_l\nonumber\\
&\equiv \hat{H}_l^{0}+\hat{H}_l^{\text{int}},
\label{HamTotal}
\end{align}
where $V(\vec{r})$ is the total trapping potential of the ladder, which can be approximated by a truncated combination of all the ring potentials $V_j(r)=\frac{1}{2}m\omega^2(R-r_j)^2$, and $g$ is the strength of the $s-$wave atom-atom interactions. The kinetic part of the Hamiltonian, $\hat{H}_l^0$, describes the tunneling dynamics of the states of total OAM $l$ between neighbouring rings as well as between the two degenerate states within the same ring. This type of dynamics was studied in detail in \cite{geometricallyinduced}. By analysing the mirror symmetries of the two-ring problem, it can be shown that there are only three independent coupling amplitudes which correspond, respectively, to the tunneling within a ring, $J_1^l=\braket{j,\pm l|\hat{H}|j,\mp}$, the tunneling to a neighbouring ring without exchange of the OAM circulation, $J_2^l=\braket{j,\pm l|\hat{H}|j+1,\pm l}$, and the tunneling to a neighbouring ring with exchange of the OAM circulation, $J_3^l=\braket{j,\pm l|\hat{H}|j+1,\mp l}$. Assuming that the central angle takes values $\Theta >\pi/3$, one can consider that only nearest-neighbour sites are coupled. By making this approximation and choosing the origin of phases to be along the line that unites the sites $2i\leftrightarrow 2i+1$, so that $\varphi_0=\pi-\Theta$ along the $2i-1\leftrightarrow 2i$ direction, the single particle terms of the Hamiltonian take the form
\begin{align}
\hat{H}_l^0&=J_1^l\sum_{j=1}^{N}\hat{a}_{+l}^{j\dagger}\hat{a}_{-l}^{j}(1+e^{-i2l\Theta})\nonumber\\
&+ J_2^l\sum_{i=1}^{N/2}\hat{a}_{+l}^{2i\dagger}(\hat{a}_{+l}^{2i+1}+\hat{a}_{+l}^{2i-1})+\hat{a}_{-l}^{2i\dagger}(\hat{a}_{-l}^{2i+1}+\hat{a}_{-l}^{2i-1})\nonumber\\
&+J_3^l\sum_{i=1}^{N/2}\hat{a}_{+l}^{2i\dagger}(\hat{a}_{-l}^{2i+1}+e^{-i2l\Theta}\hat{a}_{-l}^{2i-1})\nonumber\\
&+J_3^l\sum_{i=1}^{N/2}\hat{a}_{-l}^{2i\dagger}(\hat{a}_{+l}^{2i+1}+e^{i2l\Theta}\hat{a}_{+l}^{2i-1})+\text{h.c.}
\label{Hoperators}
\end{align} 
Assuming that only on-site interactions take place, the interacting part of the Hamiltonian can be written as
\begin{equation}
\hat{H}_l^{\text{int}}=\frac{U}{2}\sum_{j=1}^{N} \hat{n}_{+l}^j(\hat{n}_{+l}^j-1)+\hat{n}_{-l}^j(\hat{n}_{-l}^j-1)+4\hat{n}_{+l}^j\hat{n}_{-l}^j,
\end{equation}
with $U=g\int d\vec{r}|\psi_0(r)|^4$.

We now focus on the scenario in which the ladder is at unit filling, $M=N$, and the interaction strength is positive and much larger than the tunneling energies, $U\gg |J_2^l|,|J_3^l|$. In this particular situation, the system is in a Mott Insulator phase, in which the most energetically favoured states are those where all rings are occupied by a single boson. Due to the OAM degree of freedom, the ladder has $2^N$ such states, which correspond to all possible configurations of singly occupied rings with positive or negative OAM circulation of the boson. We can perform a direct mapping between these states and a spin-$1/2$ configuration by identifying a spin up (down) for each ring with a boson in the state of positive (negative) OAM circulation, i.e., $\ket{j,+l}\rightarrow \ket{\uparrow}_j; \ket{j,-l}\rightarrow \ket{\downarrow}_j$. Furthermore, we can define the spin-flip operators $\sigma_j^{\pm}=a_{\pm l}^{j\dagger}a_{\mp}^{j}$, which can be expressed in terms of the $x$ and $y$ Pauli matrices as $\sigma_j^{\pm}=\frac{1}{2}(\sigma_j^x\pm i\sigma_j^y)$. We also define the $z$ Pauli matrix as $\sigma_j^z=a_{+l}^{j\dagger}a_{+l}^{j}-a_{-l}^{j\dagger}a_{-l}^{j}$ and the spin up and down projectors $P_j^{\uparrow}=a_{+l}^{j\dagger}a_{+l}^{j}$, $P_j^{\downarrow}=a_{-l}^{j\dagger}a_{-l}^{j}$. 

\begin{figure}[t!]
\centering
\includegraphics[width=\linewidth]{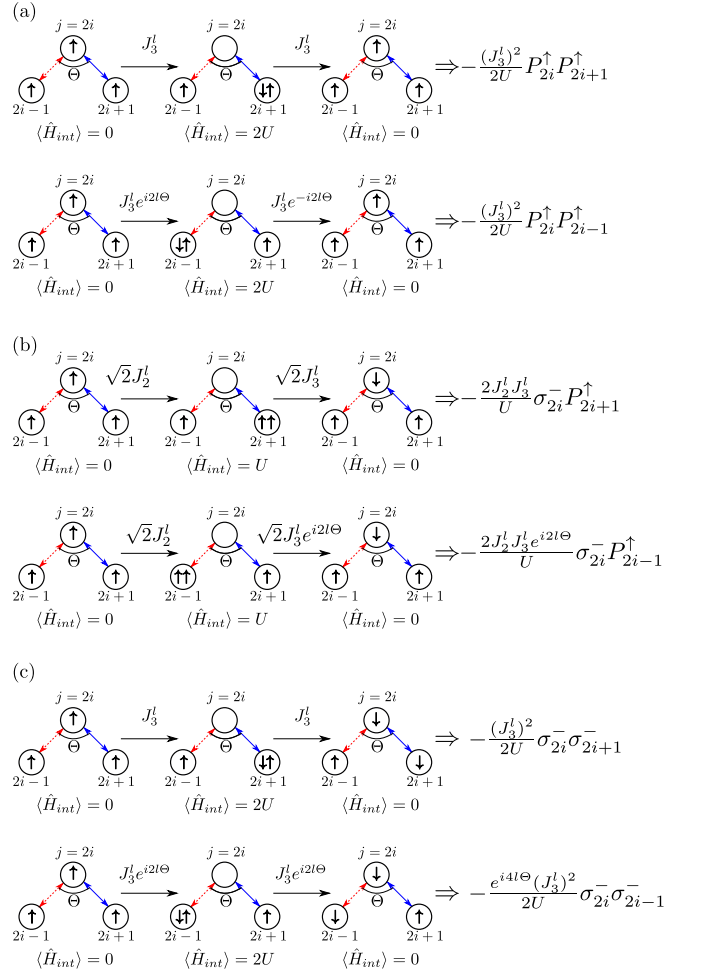}
\caption{Sketch of some of the second order processes that take place in the Mott insulator regime and their associated amplitudes. (a) Second order processes not involving any flipping of the spins. (b) Second order processes leading to the flipping of one spin at site $2i$. (c) Second order processes leading to the simultaneous flipping of the spins at sites $2i$ and $2i+1$ (upper figure) and at sites $2i$ and $2i-1$ (lower figure).}
\label{2ndordproc}
\end{figure}

The physics of the ladder in the Mott Insulator phase can be described by an effective model that incorporates interaction terms between the neighbouring spins induced by the kinetic part of the Hamiltonian, $\hat{H}_l^0$, which we treat as a perturbation. This follows the same form as the usual reduction of Hubbard, Bose-Hubbard and related models in the Mott Insulator regime to spin models. More details on the derivation of the effective model can be found in Appendix \ref{EffHam}. The new element here comes from the tunnelling phase, which can lead to non-trivial dependence of the effective model on the geometry of the lattice. The resulting effective Hamiltonian contains four types of processes. As sketched in Fig.~\ref{2ndordproc}, there are three different kinds of second order processes induced by the effective interaction: those in which the final states of the rings $j$ and $j\pm 1$ have 0, 1, or 2 spins flipped with respect to the initial state. Furthermore, there is a first order process which corresponds to the flipping of a single spin due to the self-coupling. Next, we compute separately the amplitudes corresponding to each of these different processes.
\subsection*{Processes involving no spin flips}
The first type of second order processes that we consider are those in which the initial and final states coincide, i.e., no spins are flipped. In Fig. \ref{2ndordproc}(a) we show two examples of such processes, one in which a boson at ring $j=2i$ tunnels to $j=2i+1$ and back and another one in which it tunnels to $j=2i-1$. In spite of the fact that along the direction $2i\leftrightarrow 2i-1$ there are hopping phases in the tunneling terms that exchange angular momentum, in the total second order processes they cancel out because there have to be two opposite flips in order to come back to the initial state. Thus, the total amplitude of these processes is the same regardless of the direction of the interaction. For each pair of interacting rings, there are in total 16 different second order processes not involving any total spin flip, which correspond to the 4 possible two-ring spin configurations and the 4 possible doubly occupied virtual states that mediate the interaction. Adding up all the amplitudes of these processes and using the spin notation, we find that the part of the effective Hamiltonian corresponding to these processes reads
\begin{equation}
\hat{H}_{j\leftrightarrow j\pm 1}^{0 \text{flip}}=-\frac{3((J_2^l)^2-(J_3^l)^2)}{2U}\sigma_{j}^z\sigma_{j\pm 1}^z-\frac{5((J_2^l)^2+(J_3^l)^2)}{2U}\mathbb{I}
\label{Heff0flips}
\end{equation}
We note that these amplitudes do not depend on the position of the ring $j$ inside the unit cell where it belongs.   
\subsection*{Processes involving one spin flip}
The first possibility to flip a single spin in the ring $j$ is by the action of the self-coupling $J_1^l$. The total amplitude for this process is $(1+e^{i2l\Theta})\sigma_j^-+(1+e^{i2l\Theta})\sigma_j^+=J_1^l(\sigma_j^x(1+\cos 2l\Theta)+\sigma_j^y\sin 2l\Theta)$.

Additionally, a single spin can be flipped by means of second order processes. In Fig. \ref{2ndordproc} (b) we show two examples of second order processes that lead to the flipping of a spin at the ring $j=2i$, one with a virtual interaction occurring at $j=2i+1$ and another one mediated by the ring $j=2i-1$. In this case, the amplitudes of the processes depend on the direction of the interaction: when they occur along $j=2i\leftrightarrow 2i+1$ they are real, whereas along the line $j=2i\leftrightarrow 2i-1$ a net hopping phase appears. Adding up all the amplitudes of the 12 different second order processes that lead to the flipping of a single spin, we find
\begin{equation}
\hat{H}_{j}^{1 \text{flip}}=\left(J_1^l-\frac{3J_2^lJ_3^l}{U}\right)(\sigma_j^x(1+\cos 2l\Theta)+\sigma_j^y\sin 2l\Theta)
\label{Heff1flips}
\end{equation}
Again, the total amplitude does not depend on the position of $j$ inside the unit cell because all rings are coupled to a ring along each of the two directions with different hopping phases. For values of the central angle such that $l\Theta=\pi/2$ ($\text{mod} 2\pi$), the single spin-flip amplitude vanishes, and for $l\Theta=0,\pi$ ($\text{mod}  2\pi$), it becomes $\hat{H}_{j}^{1 \text{flip}}=\left(2J_1^l-\frac{6J_2^lJ_3^l}{U}\right) \sigma_j^x$.
\subsection*{Processes involving two spin flips}
Finally, in Fig. \ref{2ndordproc} (c) we show two examples of second order processes that lead to the simultaneous flipping of two spins. As in the case of single spin-flip processes, along the $j=2i\leftrightarrow 2i-1$ interaction direction there are no total cancellations of the hopping phases. Thus, the sum of the amplitudes of these processes depends on the direction along which the bosons interact. Adding up the 8 possible processes that lead to the simultaneous flipping of two spins in the final states, we find
\begin{equation}
\hat{H}_{2i\leftrightarrow 2i+1}^{2 \text{flip}}=-\frac{(J_2^l)^2+(J_3^l)^2}{2U}\sigma_{2i}^x\sigma_{2i+1}^x-\frac{(J_2^l)^2-(J_3^l)^2}{2U}\sigma_{2i}^y\sigma_{2i+1}^y
\label{Heff2flips_real}
\end{equation} 
\begin{align}
\hat{H}_{2i\leftrightarrow 2i-1}^{2 \text{flip}}=&-\frac{(J_2^l)^2+\cos 4l\Theta(J_3^l)^2}{2U}\sigma_{2i}^x\sigma_{2i-1}^x\nonumber\\
&-\frac{(J_2^l)^2-\cos 4l\Theta(J_3^l)^2}{2U}\sigma_{2i}^y\sigma_{2i-1}^y\nonumber\\
&-\frac{\cos 4l\Theta(J_3^l)^2}{2U}(\sigma_{2i}^x\sigma_{2i-1}^y+\sigma_{2i}^y\sigma_{2i-1}^x)
\label{Heff2flips_complex}
\end{align}  
For central angles such that $l\Theta=\pi/2,\pi/4$ ($\text{mod}2\pi$), the two-spin flip processes have equal amplitude along the two directions.
\section{XYZ models}
\label{models}
By tuning the central angle $\Theta$, the amplitudes of the second order processes can be modified in order to engineer a range of quantum magnetic models. Next, we give examples of specific geometric arrangements of the ring potential ladder that lead to interesting effective spin-$1/2$ models.
\subsection{XYZ model without external field}
For central angles $\Theta_s^l=(2s+1)\pi/2l$, with $s\in \mathbb{N}$, the single-spin flip term vanishes and the two-spin flip term becomes isotropic. Summing over all the sites and processes and neglecting the constant term that appears in the zero-spin flip terms, we arrive at the following effective Hamiltonian of the Mott insulator regime 
\begin{equation}
\hat{H}_{\text{eff}}^l(\Theta_s^l)=\sum_{j=1}^{N}J_{xx}^l\sigma_{j}^{x}\sigma_{j+1}^{x}
+J_{yy}^l\sigma_{j}^{y}\sigma_{j+1}^{y}+J_{zz}^l\sigma_{j}^{z}\sigma_{j+1}^{z},
\label{HeffXYZ}
\end{equation}
where $J_{xx}^l=-((J_2^l)^2+(J_3^l)^2)/2U$, $J_{yy}^l=-((J_2^l)^2-(J_3^l)^2)/2U$ and $J_{zz}^l=-3((J_2^l)^2-(J_3^l)^2)/2U$. The Hamiltonian \eqref{HeffXYZ} is equivalent to the one of the Heisenberg $XYZ$ model, which is a prominent model of quantum magnetism and is exactly solvable \cite{BaxterBook}.
\subsection{XYZ model with external field}
For values of the central angle $\bar{\Theta}_s^l=2\Theta_s^l$, the single-spin flip amplitude contains only $\sigma_x$ one-body terms and the two-spin flip term remains isotropic. Thus, for these particular values of $\Theta$ the effective model of the ladder becomes
\begin{align}
\hat{H}_{\text{eff}}^l(\Theta_s^l)=&\sum_{j=1}^{N}J_{xx}^l\sigma_{j}^{x}\sigma_{j+1}^{x}
+J_{yy}^l\sigma_{j}^{y}\sigma_{j+1}^{y}+J_{zz}^l\sigma_{j}^{z}\sigma_{j+1}^{z}\nonumber\\
&+h^l\sum_{j=1}^{N}\sigma_{j}^{x},
\label{HeffXYZ_field}
\end{align}
with $h^l=2J_1^l-6J_2^lJ_3^l/U$. The Hamiltonian \eqref{HeffXYZ_field} corresponds to a $XYZ$ Heisenberg model with an external field $h^l$ along the $x$ direction. In the system of $p-$ orbital bosons described in \cite{XYZpinheiro}, the external magnetic field is created by the imbalance between the $p_x$ and $p_y$ interaction strengths and on-site energies, while in the ladder of rings loaded with OAM states that we consider here it arises as a consequence of the geometry of the system. 
\subsection{XYZ model with staggered fields}
\begin{figure}[h!]
\centering
\includegraphics[width=\linewidth]{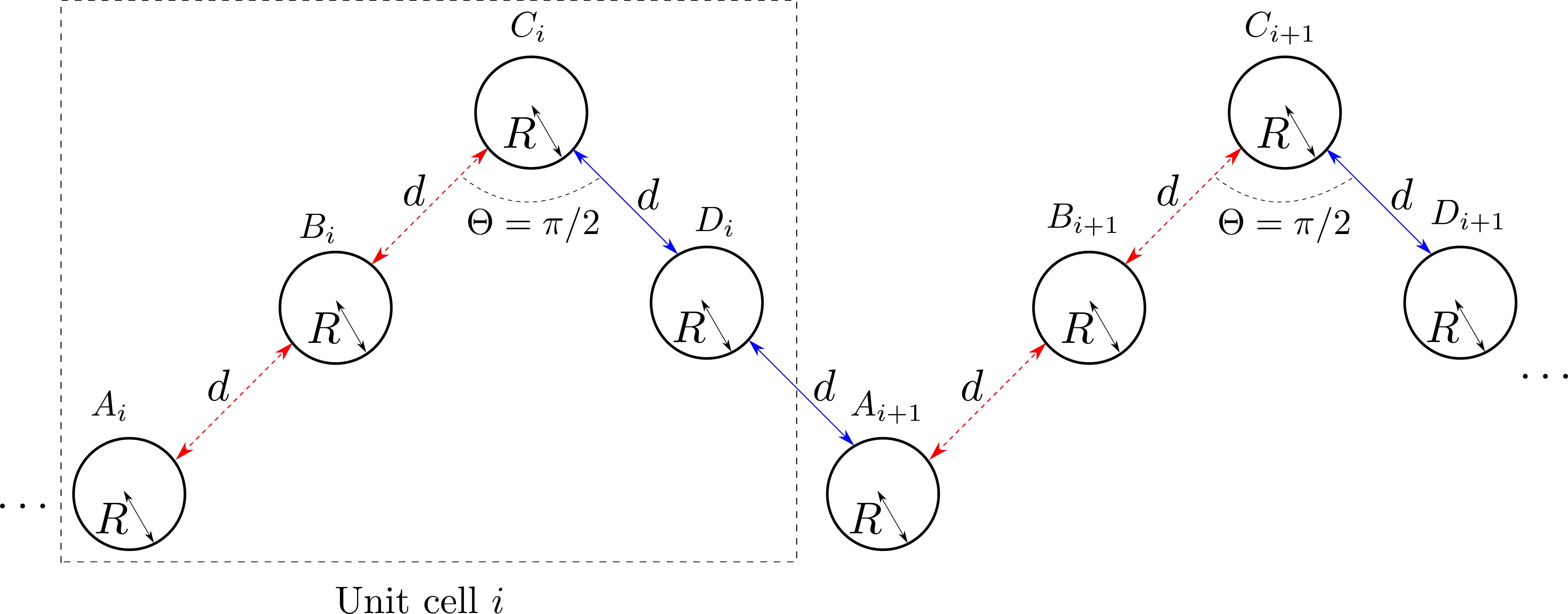}
\caption{Ladder of ring potentials with four sites per unit cell. The origin of phases is taken along the direction $A_i\leftrightarrow B_i \leftrightarrow C_i$ (indicated with blue straight arrows), so that the couplings are taken real along this direction and all hopping phases appear in the $C_i\leftrightarrow D_i \leftrightarrow A_{i+1}$ links (indicated with red dashed arrows). The distance between the closest points of two nearest-neighbour rings is $d$.}
\label{PhySystem2}
\end{figure}
By tuning the geometry of the ring potential lattice, it is also possible to obtain effective models in which the spin-$1/2$ Hamiltonian is not uniform across all sites. As an example of a system in which this can be engineered, we consider the ladder with four sites per unit cell depicted in Fig. \ref{PhySystem2} loaded with bosons in the OAM manifold $l=1$. At sites $A_i$ and $C_i$, the one-spin flip terms cancel. The $B_i$ and $D_i$ sites behave as if they belonged to a simple ladder of central angle $\pi$, but since they are coupled to rings in perpendicular directions, a relative phase will appear between them. Choosing the origin of phases along the line $A_i\leftrightarrow B_i\leftrightarrow C_i$, the effective spin Hamiltonian of this system reads 
\begin{align}
H_{\text{eff}}^{l=1}(\Theta=\pi/2)&=
\sum_{j}J_{xx}^1\sigma_{j}^{x}\sigma_{j+1}^{x}+J_{yy}^1\sigma_{j}^{y}\sigma_{j+1}^{y}
+J_{zz}^1\sigma_{j}^{z}\sigma_{j+1}^{z}\nonumber\\
&+h^1\sum_{i}\sigma_{B_i}^{x}-\sigma_{D_i}^{x}.
\label{Heff_general_longer_ladder_pi/2}
\end{align}
In the model \eqref{Heff_general_longer_ladder_pi/2}, the external magnetic fields appear in a staggered pattern only at the $B_i$ and $D_i$ sites.
\section{Control over parameters and quantum phases through the trap geometry}
\label{control}
\begin{figure}[t!]
\centering
\includegraphics[width=\linewidth]{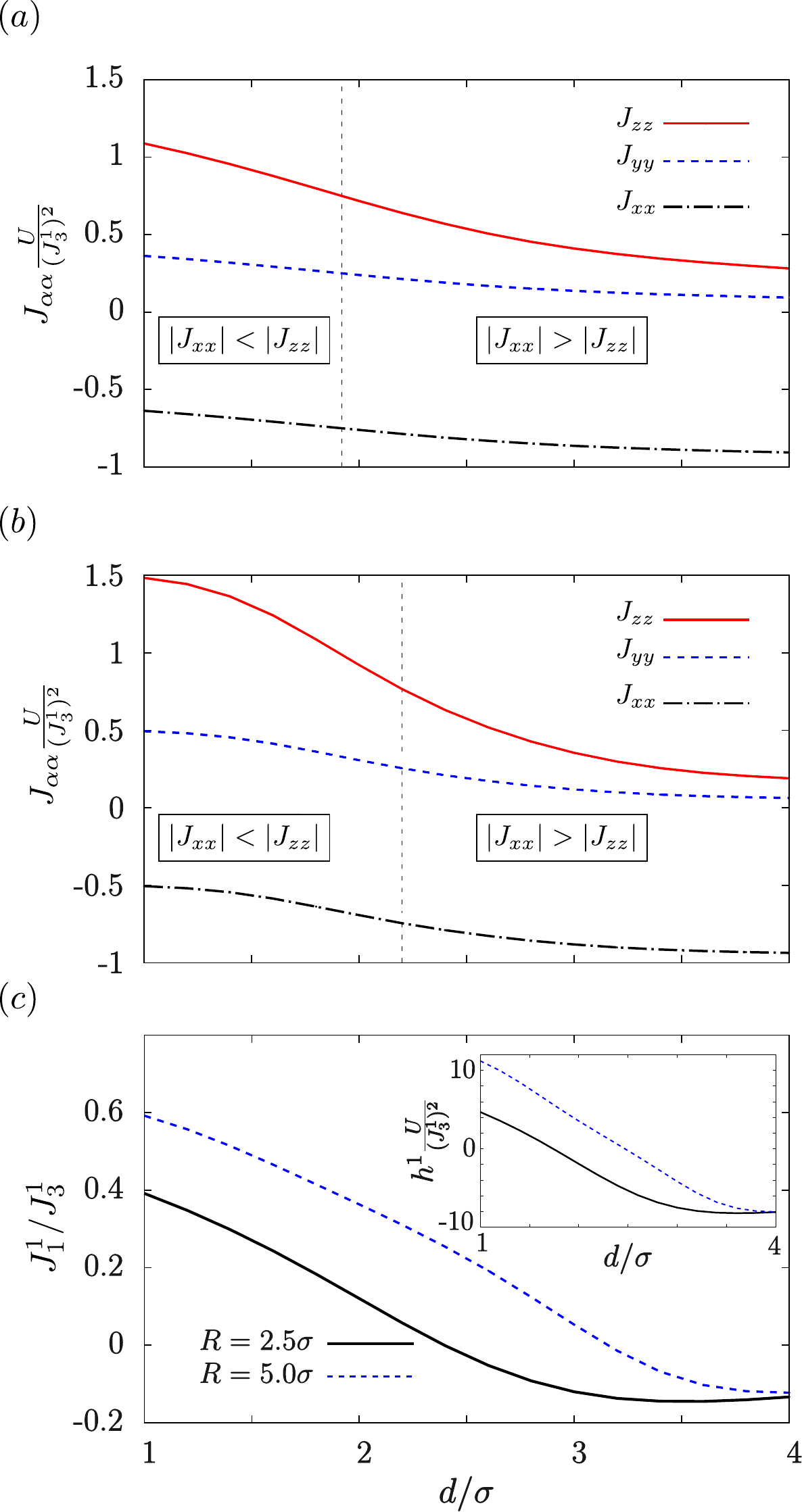}
\caption{Dependence of the effective couplings of the $XYZ$ model \eqref{HeffXYZ} on the inter-ring separation $d$ for rings of $R=2.5\sigma$ (a) and rings of $R=5.0\sigma$ (b). The dashed vertical lines mark the value of $d$ for which the transition of the $XYZ$ model without external field \eqref{HeffXYZ} between the $z$-antiferromagnet and the $x$-ferromagnet occurs. (c) Dependence of the ratio $J_1^1/J_3^1$ on the inter-ring separation. The inset shows the dependence of $h^1$ on the inter-ring separation taking $U/J_3^1=20$ for all values of $d$.}
\label{effective_couplings_distance}
\end{figure}
In this section, we describe how the effective parameters of the spin$-1/2$ models can be tuned by modifying the size of the ring potentials and the separation between them. We also discuss the different phases of the $XYZ$ model without external field \eqref{HeffXYZ} that can be explored with the system and we analyze their robustness against deviations of the central angle of the ladder from the values $\Theta_s^l$ that yield the effective model \eqref{HeffXYZ}. Although we focus on the case of $l=1$ OAM states, our considerations can be generalized to other OAM manifolds in a straightforward manner. 
\subsection{Control of the effective model parameters}
In a two-ring system, it is possible to compute the values of the tunneling parameters $\{J_1^1,J_2^1,J_3^1\}$ by calculating numerically the energies of the OAM eigenstates of the system, which are related to the hopping strengths via a four-state model \cite{geometricallyinduced}. This procedure determines the dependence of the relative values of the couplings on the ring radius $R$ and the separation between rings $d$ \cite{diamondchain}. For small values of the inter-ring distance $d\sim \sigma$, $J_3^1$ is several times larger than $J_2^1$. As shown in Fig. \ref{effective_couplings_distance} (b), which corresponds to a ring of $R=5\sigma$, in the most extreme limit of this regime the couplings of the effective model fulfill the relation $J_{xx}\approx -J_{yy}=-J_{zz}/3$. For rings of smaller radius, as the ring of $R=2.5\sigma$ corresponding to Fig. \ref{effective_couplings_distance} (a), $J_2^1$ and $J_3^1$ are more similar at small values of $d$, and therefore the ratio $|J_{zz}|/|J_{xx}|$ is smaller. However, for both values of $R$ there is a range of inter-ring separations for which the condition $3J_{yy}=J_{zz}>-J_{xx}$ holds. 

In this parameter regime, the $XYZ$ model without external field \eqref{HeffXYZ} is in a anti-ferromagnetic phase in the $z$ direction \cite{PhaseDiagram}. As $d$ is increased, $J_3^1$ and $J_2^1$ become more similar, until the critical point $J_{zz}=-J_{xx}$ is reached. This point, which is signalled with dashed vertical lines in Fig. \ref{effective_couplings_distance} (a), (b), marks the transition to a ferromagnetic phase in the $x$ direction \cite{PhaseDiagram}. In the limit of very large $d$, $J_3^1=J_2^1$ and therefore $J_{zz}=J_{yy}=0$. 

The behaviour of the ratio $J_1^1/J_3^1$ as a function of $d$ is shown in Fig. \ref{effective_couplings_distance} (c) for rings of $R=2.5\sigma$ and $R=5\sigma$. For small values of $d$, $J_1^1$ has the same sign as $J_3^1$ and is of the same order or higher. As $d$ is increased, $J_1^1/J_3^1$ decreases until zero, and then it remains small and negative. As shown in the inset of Fig. \ref{effective_couplings_distance} (c), this behaviour of the $J_1^1/J_3^1$ ratio translates into the effective field $h^1$ being positive at small values of $d$, and as $d$ is increased decreasing to a minimum negative value and finally remaining negative and with an approximately constant value.
\begin{figure}[t!]
\centering
\includegraphics[width=0.9\linewidth]{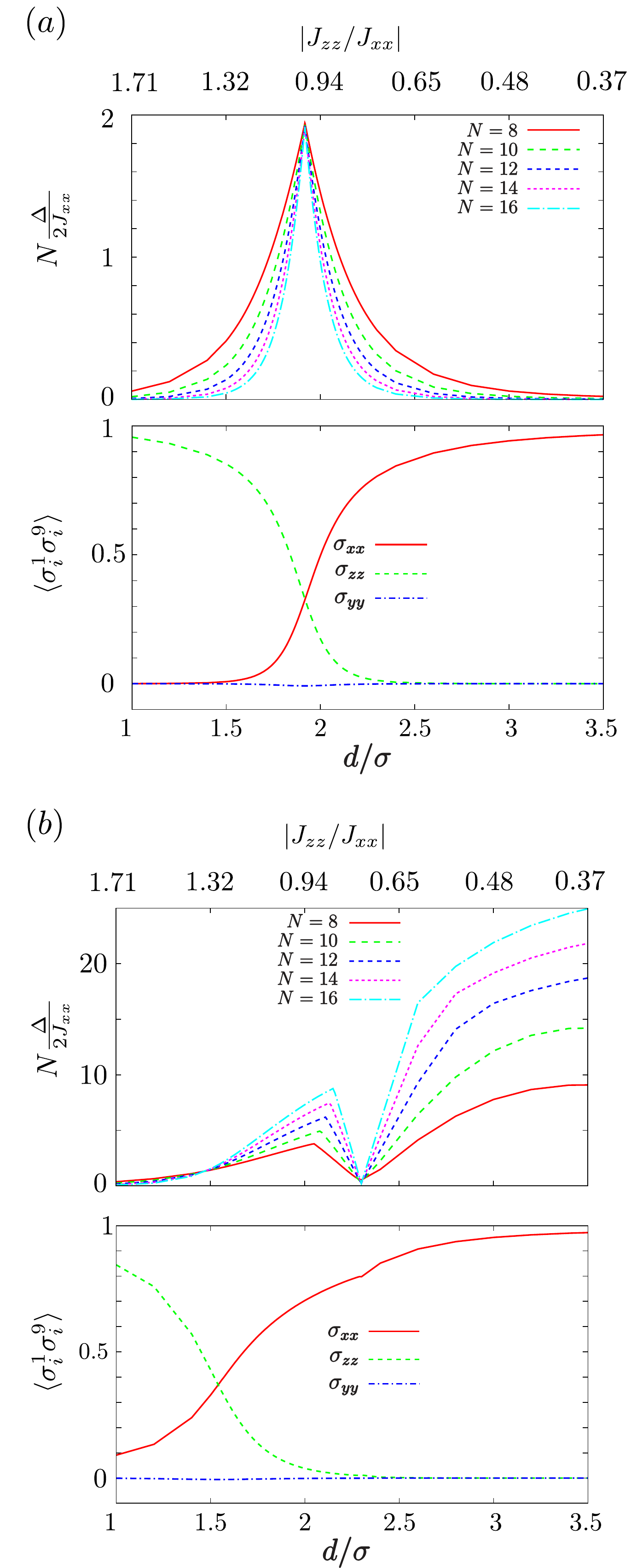}
\caption{Upper plots in (a) and (b): dependence of the energy difference between the ground and first excited states on the inter-ring distance $d$ for ladders of different sizes formed by rings of radius $R=2.5\sigma$. Lower plots in (a) and (b): correlations between spin 1 and 9 in a ladder of $N=16$ spins with PBC formed by rings of $R=2.5\sigma$. In (a) the central angle of the ladder is $\Theta=\frac{\pi}{2}$ and in (b) $\Theta=0.48\pi$.}
\label{phasetransition}
\end{figure}
\subsection{Example properties of the obtainable quantum phases}
In order to analyse numerically the phases of ladders with different central angles $\Theta$, we have performed exact diagonalization in chains of up to $N=16$ spins with Periodic Boundary Conditions (PBC). If a quantum critical point exists, we expect that the energy gap $\Delta$ between the ground and first excited state scales with the system size as $\Delta\sim \frac{1}{N}$ \cite{MottEfield}. Therefore, we have searched for the critical point by plotting for ladders of different sizes the quantity $\Delta N$ as a function of the inter-ring separation $d$ and looking at the point where all the lines intersect. In order to confirm directly the presence of the transition point between the $z-$antiferromagnetic and the $x-$ferromagnetic phases, we have also computed for a ladder of $N=16$ spins with PBC the ground state correlations between two fixed spins as a function of $d$.

In Fig. \ref{phasetransition} (a) we show the results of these two analyses for a ladder formed by rings of $R=2.5\sigma$, filled with bosons in $l=1$ OAM states and with a central angle $\Theta=\frac{\pi}{2}$, which is described by the $XYZ$ model without external field \eqref{HeffXYZ}. The upper plots shows the dependence of $\Delta N$ on $d$ for different system sizes. As expected, all the lines intersect at the value of $d$ for which $|J_{zz}|=|J_{xx}|$, where the phase transition occurs. As shown in the lower plot, in the $z-$antiferromagnetic phase, the $zz$ correlation is higher than the $xx$ one. As $d$ increases, the $zz$ correlation decays and the $xx$ one increases, until they reach the same value at an inter-ring distance that coincides with the corresponding one for the critical point.

In Fig. \ref{phasetransition} (b), we perform the same analysis for a ladder as in Fig. \ref{phasetransition} (a) but with a central angle $\Theta=0.48\pi$. In the upper plot, we observe that there are two points where the $\Delta N$ lines intersect. The one that occurs for a smaller value of $d$ corresponds to a point where the $zz$ and $xx$ correlations become equal. Therefore, it marks the transition between the $z-$antiferromagnetic and the $x-$ferromagnetic phases. This transition occurs at a smaller value of $d$ than in the ladder with a central angle $\Theta=\frac{\pi}{2}$ because of the presence of the magnetic field along the $x$ direction. For values of the central angle more deviated from $\frac{\pi}{2}$, the presence of the magnetic field destroys the $z-$antiferromagnetic phase and the transition does not occur. The other point where the $\Delta N$ curves intersect occurs at a longer value of $d$ which coincides with the point where the external magnetic field vanishes (see the inset of Fig.\ref{effective_couplings_distance} (b)). Therefore, it corresponds to a global change of orientation of the spins in the $x-$ferromagnetic phase.
\section{Considerations for experimental implementations}
\label{experimental}

\subsection{Realization of the ring lattices}
\label{lattices}
The arrays of ring traps considered in the previous sections could be created by means of several different techniques. Since they were first proposed a few years ago \cite{ring4,TAAP}, time-averaged adiabatic potentials have proven to be a powerful tool to trap ultracold atoms in on-demand potential landscapes \cite{ring5,TimeAveragedBECRing,TimeAveragedBECRing2}. Recently, it has also been shown that digital micro-mirror devices allow to create trapping potentials with arbitrary shapes \cite{PatterningDMD}, and in particular a double ring trap has been realized \cite{DoubleRing}. Both of these already demonstrated approaches could be adapted to create lattices of ring potentials. Conical refraction, which is a phenomenon that occurs when a focused light beam passes along an optic axis of a biaxial crystal, has also been used to trap ultracold atoms in ring geometries \cite{ring7}. With this technique, arrays of ring potentials could be generated by reproducing with microlenses the intensity pattern of a laser beam traversing a single crystal. Alternatively, the combination of split lenses and spatial light modulators \cite{PatterningLens} could also be used to implement arrays of light rings with any desired geometry. 
\subsection{Experimental readout of the spin states}
\label{readout}
Making use of a scheme of two-photon stimulated Raman transitions in the Lamb-Dicke regime analogous to the one discussed in \cite{XYZpinheiro}, the OAM states of a single site could be addressed separately. By tuning the frequencies of the lasers, side-band transitions between the ground state of the ring potential and the $\pm l$ OAM states could be induced. Once the OAM states are encoded in the internal atomic states, a quantum gas microscope \cite{QGasMicro1,QGasMicro2} could be used to read the spin states of the effective models with single-site resolution. 

In order to select a specific OAM mode, two different approaches could be followed. As pointed out in \cite{XYZpinheiro}, one option would be to choose the laser beams such that their wave vector difference is oriented along the $x$ or $y$ direction. In that manner, the lasers would only interact with states that have nodes along the $x$ or $y$ axes, which can be expressed as symmetric or anti-symmetric superpositions of OAM modes. Alternatively, with the aid of
e.g. spatial light modulators, one could also make a small
adiabatic deformation of the ring trap in order to break cylindrical symmetry and  induce
an energy splitting between the dressed OAM states such that they can be independently resolved.        
\subsection{Collisional stability of the OAM states}
\label{stability}
An important question concerning the feasibility of the experimental realization of the system is whether collisional processes may cause transitions between states with different OAM that invalidate the assumption that all the atoms remain always in the same OAM manifold \cite{collisions}. These collisional processes are described in the Born approximation by the operator
\begin{equation}
\hat{U}=\frac{g}{2}\int d\vec{r} \hat{\Psi}^{\dagger}\hat{\Psi}^{\dagger}\hat{\Psi}\hat{\Psi},
\label{CollisionsBorn}
\end{equation}
where $\hat{\Psi}=\sum_{l=0}^{\infty}\hat{\Psi}_l$ is the full bosonic field operator of the lattice, given by the sum of all the field operators of the different OAM manifolds \eqref{bosonicfield}. In order to analyse the stability of the OAM states under the most relevant collisional processes, it is enough to restrict ourselves to the examination of two-boson states in a single ring. This is because the collisional interactions are strongly dominated by on-site processes. In the strongly interacting regime, the role of tunnelling is significantly reduced. In the opposite limit where atoms are delocalised, we would have to consider the full Bloch Band widths for energy conservation in a collisional process taking atoms to other OAM states.

The operator \eqref{CollisionsBorn} only yields non-zero matrix elements between states with the same total OAM. Since the separation between the OAM energy levels is anharmonic (see eq. \eqref{energiesOAM}), the allowed transitions between two-boson states that are not within the same OAM manifold are always off-resonant. For rings of radius of the order of a few $\sigma$, the smallest possible energy difference between the states, which is of the order of $E_c$, is one order of magnitude bigger than the transition matrix element between them, which is of the order of $U$. Therefore, the system is not destabilized by collisions that take two atoms in a given OAM manifold to other OAM states. Moreover, since we are assuming that the gas of ultracold atoms is in the Mott insulator phase at unit filling, the ground state has only very small contributions from states with more than one atom per site. Therefore, the occurrence of these collisional processes is suppressed in the first place by the population distribution of the many-body states in the ground state \cite{pbandexperiment}.
\section{Conclusions}
We have shown that ultracold bosons carrying OAM in arrays of cylindrically symmetric potentials realize a variety of spin$-1/2$ models of quantum magnetism in the Mott insulator regime at unit filling. By means of second-order perturbation theory, we have computed explicitly the dependence of the effective couplings on the relative angle between the traps and demonstrated that several models of interest such as the $XYZ$ model with uniform or staggered external fields can be obtained. We have discussed how the relative strength of the effective coupling parameters can be tuned and which phases of the $XYZ$ model without external field can be observed in a realistic set-up by performing this tuning. Furthermore, we have analysed the effect of small changes of the relative angles between the traps on these phases. We have also discussed single-site addressing techniques that allow to retrieve the state of each individual spin. Finally, we have analysed the collisional stability of the system and concluded that the anharmonic energy spacing between OAM states introduced by the ring geometry extends the lifetime of the Mott state. Therefore, the scheme presented in this work serves as a versatile toolbox for the quantum simulation of magnetic models and has a feasible experimental implementation. 
\label{conclusions}
\acknowledgements
G.P., J.M., and V.A. gratefully acknowledge financial support  from  the  Ministerio  de  Econom\'ia  y  Competitividad, MINECO, (FIS2014-57460-P,  FIS2017-86530-P) and from the Generalitat de Catalunya (SGR2017-1646). G.P. acknowledges  financial  support  from  MINECO  through  Grant  No. BES-2015-073772. Work at the University of Strathclyde was supported by the EPSRC Programme Grant DesOEQ (EP/P009565/1). We thank Anton Buyskikh and Luca Tagliacozzo for helpful discussions.
\appendix
\section{Derivation of the effective Hamiltonian}
\label{EffHam}
To derive the effective Hamiltonian, we define a projector $\hat{M}$ to the Mott space of singly occupied sites as well as the projector to the space orthogonal to this one $\hat{O}=1-\hat{M}$. In terms of these operators, the Schr\"odinger equation $(\hat{H}_l^{0}+\hat{H}_l^{\text{int}})\ket{\Psi}=E\ket{\Psi}$ can be decomposed as \cite{XYZpinheiro}
\begin{align}
&(\hat{O}\hat{H}_l^{0}\hat{O}+\hat{O}\hat{H}_l^{0}\hat{M}
+\hat{O}\hat{H}_l^{\text{int}}\hat{O}+\hat{O}\hat{H}_l^{\text{int}}\hat{M})\ket{\Psi}
=E\hat{O}\ket{\Psi}
\label{SchrO}\\
&(\hat{M}\hat{H}_l^{0}\hat{O}+\hat{M}\hat{H}_l^{0}\hat{M}
+\hat{M}\hat{H}_l^{\text{int}}\hat{O}+\hat{M}\hat{H}_l^{\text{int}}\hat{M})\ket{\Psi}
=E\hat{M}\ket{\Psi}.
\label{SchrM}
\end{align}
The terms $\hat{M}\hat{H}_l^{\text{int}}\hat{M}$, $\hat{O}\hat{H}_l^{\text{int}}\hat{M}$ and $\hat{M}\hat{H}_l^{\text{int}}\hat{O}$ are all identically zero: the first two ones for computing two-body interactions in single-ocupied rings and the last one for computing overlaps between orthogonal spaces. Taking this fact into account, we can combine eqs. \eqref{SchrO} and \eqref{SchrM} to write  
\begin{equation}
\hat{H}_{\text{eff}}\hat{M}\ket{\Psi}=E\hat{M}\ket{\Psi},
\end{equation}
where the effective Hamiltonian reads
\begin{equation}
\hat{H}_{\text{eff}}=-\hat{M}\hat{H_l^0}\hat{O}\frac{1}{\hat{O}\hat{H}_l^{\text{int}}\hat{O}-E}\hat{O}\hat{H}_l^0\hat{M}+\hat{M}\hat{H_l^0}\hat{M}.
\label{Heff}
\end{equation}
The physical action of the first term of the effective Hamiltonian is to connect a Mott state to a state of the orthogonal space through the tunneling term of the original Hamiltonian, associate an energy to this state in the orthogonal space according to $(\hat{O}\hat{H}_l^{\text{int}}\hat{O}-E)^{-1}$ and then take the state back to the Mott subspace. All the second order processes induced by this term occur via intermediate states in which all the rings are singly occupied except for one ring, say $j$, that is empty, and the ring $j\pm 1$, that is doubly occupied. Therefore, we restrict the orthogonal subspace to the these states, which can be compactly represented by the three possible two-spin states for the doubly occupied rings, namely $\{\ket{\uparrow\uparrow}_{j\pm 1},\ket{\downarrow\downarrow}_{j\pm 1},\ket{\uparrow\downarrow}_{j\pm 1}\}$. Furthermore, since we are in the Mott insulator regime, we can assume that $(\hat{O}\hat{H}_l^{\text{int}}\hat{O}-E)^{-1}\approx (\hat{O}\hat{H}_l^{\text{int}}\hat{O})^{-1}$. In the subspace of states where only a single ring has double occupation, this operator takes the form $(\hat{O}\hat{H}_l^{\text{int}}\hat{O})^{-1}=\text{diag}\{1/U,1/U,1/2U\}$. The second term of the effective Hamiltonian, $\hat{M}\hat{H_l^0}\hat{M}$, takes into account the first order processes that occur within the subspace of singly occupied states, which are due to the self-coupling amplitude $J_1^l$.

\end{document}